\documentclass[preprint]{aastex}
\def\uka { \raisebox{-0.5ex} {\mbox{$\stackrel{<}{\scriptstyle \sim}$}}}

\def\nnerr#1#2{{\raisebox{-0.5ex} {$\stackrel{+{#2}}{\scriptstyle-{#1}}$}}}
\shortauthors{Aharonian et al.}
\shorttitle{Mkn~501 in 1998 and 1999}
\begin{document}
\title{The TeV Energy Spectrum of Mkn~501 Measured with the 
Stereoscopic Telescope System of HEGRA during 1998 and 1999}
\author{
F.~Aharonian\altaffilmark{1},
A.~Akhperjanian\altaffilmark{7},
J.~Barrio\altaffilmark{2,3},
K.~Bernl\"ohr\altaffilmark{1},
H.~B\"orst\altaffilmark{5},
H.~Bojahr\altaffilmark{6},
O.~Bolz\altaffilmark{1},
J.~Contreras\altaffilmark{2},
J.~Cortina\altaffilmark{2},
S.~Denninghoff\altaffilmark{2},
V.~Fonseca\altaffilmark{3},
J.~Gonzalez\altaffilmark{3},
N.~G\"otting\altaffilmark{4},
G.~Heinzelmann\altaffilmark{4},
G.~Hermann\altaffilmark{1},
A.~Heusler\altaffilmark{1},
W.~Hofmann\altaffilmark{1},
D.~Horns\altaffilmark{4},
C.~Iserlohe\altaffilmark{6},
A.~Ibarra\altaffilmark{3},
I.~Jung\altaffilmark{1},
R.~Kankanyan\altaffilmark{1,7},
M.~Kestel\altaffilmark{2},
J.~Kettler\altaffilmark{1},
A.~Kohnle\altaffilmark{1},
A.~Konopelko\altaffilmark{1},
H.~Kornmeyer\altaffilmark{2},
D.~Kranich\altaffilmark{2},
H.~Krawczynski\altaffilmark{1,9},
H.~Lampeitl\altaffilmark{1},
E.~Lorenz\altaffilmark{2},
F.~Lucarelli\altaffilmark{3},
N.~Magnussen\altaffilmark{6},
O.~Mang\altaffilmark{5},
H.~Meyer\altaffilmark{6},
R.~Mirzoyan\altaffilmark{2},
A.~Moralejo\altaffilmark{3},
L.~Padilla\altaffilmark{3},
M.~Panter\altaffilmark{1},
R.~Plaga\altaffilmark{2},
A.~Plyasheshnikov\altaffilmark{1,8},
J.~Prahl\altaffilmark{4},
G.~P\"uhlhofer\altaffilmark{1},
A.~R\"ohring\altaffilmark{4},
W.~Rhode\altaffilmark{6},
G.P.~Rowell\altaffilmark{1},
V.~Sahakian\altaffilmark{7},
M.~Samorski\altaffilmark{5},
M.~Schilling\altaffilmark{5},
F.~Schr\"oder\altaffilmark{6},
M.~Siems\altaffilmark{5},
W.~Stamm\altaffilmark{5},
M.~Tluczykont\altaffilmark{4},
H.~V\"olk\altaffilmark{1},
C.~Wiedner\altaffilmark{1},
W.~Wittek\altaffilmark{2}}
\altaffiltext{1}{Max Planck Institut f\"ur Kernphysik,
Postfach 103980, D-69029 Heidelberg, Germany}
\altaffiltext{2}{Max Planck Institut f\"ur Physik, F\"ohringer Ring
6, D-80805 M\"unchen, Germany}
\altaffiltext{3}{Universidad Complutense, Facultad de Ciencias
F\'{i}sicas, Ciudad Universitaria, E-28040 Madrid, Spain }
\altaffiltext{4}{Universit\"at Hamburg, II. Institut f\"ur
Experimentalphysik, Luruper Chaussee 149,
D-22761 Hamburg, Germany}
\altaffiltext{5}{Universit\"at Kiel, Institut f\"ur Experimentelle und Angewandte Physik,
Leibnizstra{\ss}e 15-19, D-24118 Kiel, Germany}
\altaffiltext{6}{Universit\"at Wuppertal, Fachbereich Physik,
Gau{\ss}str.20, D-42097 Wuppertal, Germany}
\altaffiltext{7}{Yerevan Physics Institute, Alikhanian Br. 2, 375036 Yerevan, 
Armenia}
\altaffiltext{8}{On leave from  
Altai State University, Dimitrov Street 66, 656099 Barnaul, Russia}
\altaffiltext{9}{Corresponding author: Henric Krawczynski, email: Henric.Krawczynski@mpi-hd.mpg.de}
\begin{abstract}
\vspace*{-1ex}
During 1997, the BL Lac object Mkn~501 went into an extraordinary 
state of high X-ray and TeV gamma-ray activity, lasting more than 6
months. 
In this paper we report on the TeV emission characteristics of the
source in the subsequent years of 1998 and 1999 as measured with the 
Stereoscopic 
Cherenkov Telescope System of HEGRA (La Palma, Canary Islands).
Our observations reveal a 1998--1999 mean emission level at 1~TeV of
1/3 of the flux of the Crab Nebula, a factor of 10 lower than
during the year of 1997. 
A dataset of 122 observations hours with the  HEGRA telescope system
makes it possible to assess for the first time the Mkn~501 
TeV energy spectrum for a mean flux level substantially below that of the Crab
Nebula with reasonable statistical accuracy.
Excluding the data of a strong flare, we find evidence that the
1998--1999 low-flux spectrum is substantially softer 
(by $0.44\pm0.1_{\rm stat}$ in spectral index) than the 1997 time averaged spectrum.
The 500~GeV to $\simeq$10~TeV energy spectrum can well be described by a power law model with 
exponential cutoff: ${\rm d}N/{\rm d}E \propto E^{-\alpha}\exp{(-E/E_0)}$ with
$\alpha=2.31 \pm0.22_{\rm stat}$, and $E_0=5.1 (-2.3+7.8)_{\rm stat} \, \rm TeV$.
Within statistical accuracy, also a pure power law model gives an
acceptable fit to the data: ${\rm d}N/{\rm d}E \propto E^{-\Gamma}$ with
$\Gamma=2.76\pm0.08_{\rm stat}$.
After presenting the 1998--1999~TeV characteristics of the source 
we discuss the implications of the results.
\end{abstract}
\section{Introduction}
\label{intro}
During the years of its discovery as a TeV source 
the BL Lac object Mkn~501 showed modest integral fluxes, 
during 1995 about 1/12 of the Crab flux above 300~GeV \cite{Quin:96}
and during 1996 about 1/3 of the Crab flux above 1.5~TeV \cite{Brad:97}.
In the year 1997 the source went into a state of surprisingly high activity 
in the X-ray \cite{Pian:98} and TeV energy bands 
\citep{Samu:98,Ahar:99a,Djan:99}. 
During more than 6 months the source showed a succession of very
strong flares with an average differential flux at 1~TeV approximately 
3 times higher than the flux of the Crab Nebula.

The extraordinary high TeV emission level as well as the significant
improvement of detectors and analysis methods made it possible to 
study the temporal and spectral TeV characteristics of a BL Lac 
object with unprecedented detail.
The measurements with the Stereoscopic Cherenkov Telescope System
of HEGRA revealed that while the TeV flux varied by
factors of up to 30, the TeV spectrum remained surprisingly
stable. With a typical statistical accuracy (1~$\sigma$) of the 
diurnal spectral indices between 0.1 and 0.3 
no spectral variability could be ascertained with a statistical significance 
exceeding $3\,\sigma$ \cite{Ahar:99a}.
Furthermore, dividing the data into groups according to the
flux level at 2~TeV did not reveal any evidence for a flux hardness 
correlation (statistical accuracy of the 1~TeV to 10~TeV spectral 
indices \uka~0.05).

The excellent gamma-ray statistics combined with the 20\% 
energy resolution of the HEGRA instrument resulted in the 
first detection of gamma-rays from an extragalactic 
source well beyond 10~TeV, and the first high accuracy measurement 
of an exponential cutoff in the energy region above 5~TeV,
well into the exponential regime \cite{Ahar:99b}.
From 500~GeV to $\simeq$20~TeV the differential photon spectrum could be
approximated by a power-law with an exponential cutoff:
\begin{equation}
\label{plc} 
{\rm d} N/{\rm d} E=N_0 \, (E/1\,{\rm TeV})^{-\alpha} \,
\exp{(-E/E_0)},
\end{equation} 
with $N_0=(108\pm2_{\rm stat}\pm2.1_{\rm sys})
\times 10^{-12}$ $\rm cm^{-2}\, s^{-1}\, TeV^{-1}$,   
$\alpha=1.92 \pm0.03_{\rm stat} \pm0.20_{\rm sys}$, and 
$E_0=6.2 \pm0.4_{\rm stat} (-1.5 +2.9)_{\rm sys} \, \rm TeV$.

Note that the stability of the 1997 TeV spectrum has been a matter of debate.
The CAT group reported evidence for a flux-hardness
correlation in their 1997 Mkn~501 data set: 
the spectrum of the 3 days with maximum $>250$~GeV flux in their data
set seemed to be harder than the spectrum during the rest of the 1997
observations on a statistically significant level
although they did not report an estimate of the
systematic uncertainty on this result \cite{Djan:99}.
The apparent contradiction between the HEGRA and CAT results could be
explained by the following 2 facts:
(i) Since the CAT flux-hardness correlation stems from the energy range
below 1~TeV while the HEGRA constraints on the correlation are
strongest at energies above 1~TeV, the results could indicate a
stronger flux-hardness correlation below 1~TeV than above 1~TeV \cite{Djan:99};
(ii) since the evidence for the flux-hardness correlation is based 
on the detection of a harder spectrum for a small number of
observation nights (3 nights) it could well be that the flux-hardness 
relation found for these days does not represent the behavior of the 
source during the full 1997 flaring phase.
While there are no HEGRA observations for the strongest flare in the CAT data sample
(April 16th, 1997), the HEGRA data of the day with the second highest
TeV emission in the CAT data sample (April 13th, 1997) 
also indicated a harder than average spectrum: the 1~TeV to 5~TeV
photon index of 1.87$\pm 0.14_{\rm stat}$ deviated by 2.7~$\sigma$ from the 1997 mean
photon index of 2.25 \cite{Ahar:99a}.
Therefore a harder than average spectrum for at least 2 of the 3 CAT 
nights is consistent with the HEGRA observations even if the spectrum 
hardened as much above 1~TeV as it did below 1~TeV. 
Note that the HEGRA observations of several later 1997 flares with comparable and
higher flux levels than that of April 13th did not show such a trend
of spectral hardening

In this paper we present HEGRA observations of Mkn~501 during the
years 1998 and 1999 and show how the energy spectrum evolved in the years
following the 1997 flaring phase.
After describing the telescope system as well as
the dataset in Sect.~\ref{HEGRA} we report on the 1998 and 1999 
Mkn~501 light curve and energy spectrum in Sect.~\ref{results}. 
In Sect.\ \ref{discussion} we discuss possible
implications of the results.
\section{The HEGRA Cherenkov Telescope System and the Data Sample}
\label{HEGRA}
The HEGRA collaboration operates six imaging atmospheric Cherenkov telescopes
located on the Roque de los Muchachos on the Canary island of La Palma, 
at 2200~m above sea level. 
In this paper we present the 1998--1999 Mkn 501 data taken with the 
HEGRA Cherenkov Telescopes System \cite{Kono:99a}. 
The system consists of five telescopes (CT2 - CT6) which are operated as a single
detector for the stereoscopic detection of air showers induced by 
primary gamma-rays in the atmosphere.
The telescope system has been taking data since 1996, initially with three 
and later four telescopes, 
and since fall 1998 as a complete five-telescope system. 
With an energy threshold of 500 GeV, the telescope system achieves
an angular resolution of  0.1$^\circ$ and an energy resolution of 20\%
for individual photons, as well as an approximate energy flux sensitivity 
$\nu\,$F$_\nu$ at 1 TeV of 10$^{-11}$ $\rm erg\, cm^{-2}\, s^{-1}$ (S/N = 5~$\sigma$) 
for 1 hour of observation time. 
Note that most of the data presented in this paper were taken with the
four telescope system. The telescope ``CT2'' was undergoing significant
hardware modifications. 
The results presented in the following are based on 126~hours of 
data acquired between 1998, February 28 and 1999, July 7 with Mkn~501 altitudes above 60$^\circ$.
The search of intraday flux variability makes use of
additional 27~h of data with Mkn~501 altitudes between 45$^\circ$ and 60$^\circ$.
The standard data quality criteria, analysis tools, and ``loose'' 
selection cuts for spectral studies described in detail in 
(Aharonian et al.\ 1999a-b) were applied.
\section{Experimental results}
\label{results}
\begin{figure}[bh]
\plotone{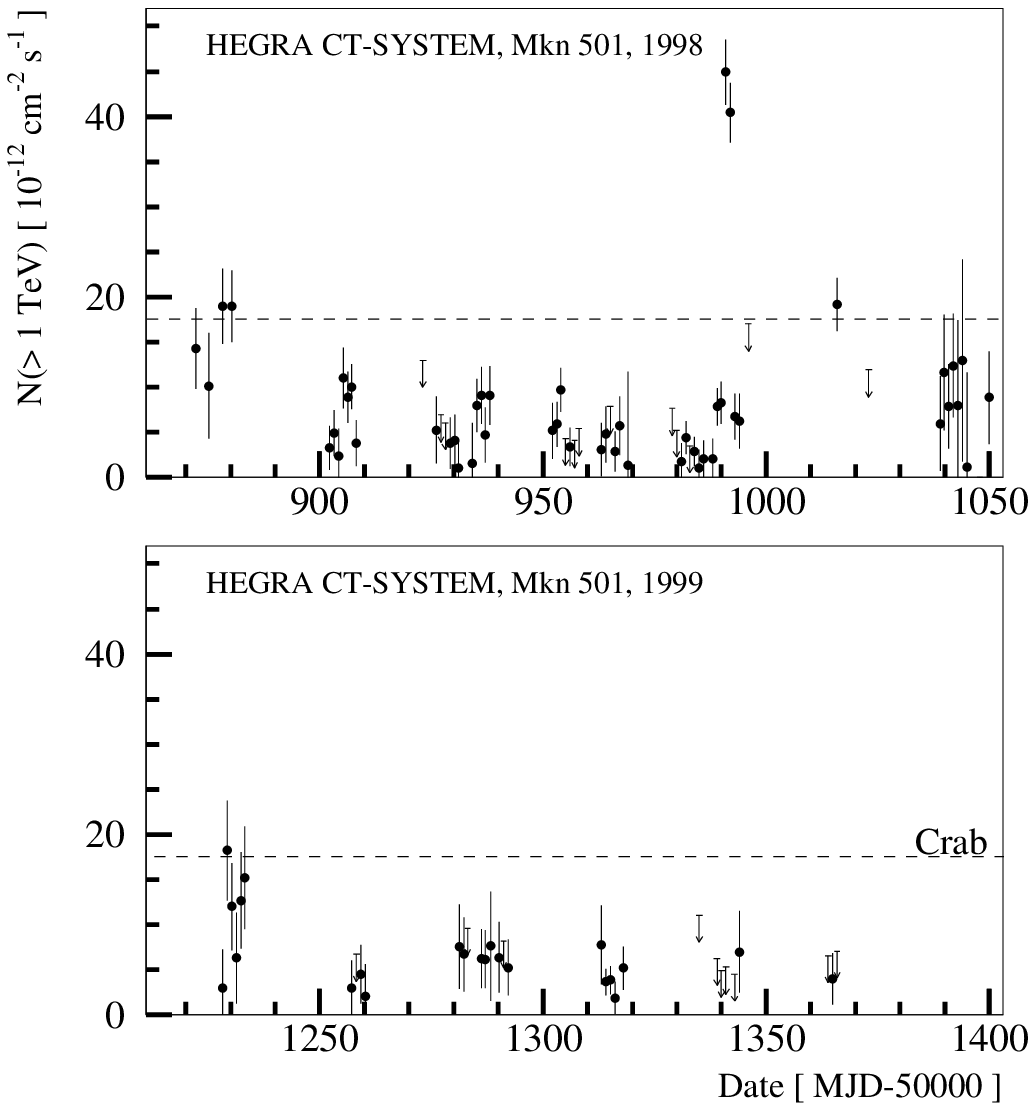}
\caption{\label{lc} \small The 1998 and 1999 light curve of Mkn~501 as measured
  with the HEGRA telescope system. The dashed line indicates the
  steady emission level of the Crab Nebula. Each data point is a
  diurnal average.
  Note that during 1997 the source reached $>1$~TeV integral flux levels
  of about 10 Crab units: HEGRA detected a maximum flux of 
  $(178\pm10_{\rm stat}) \times 10^{-12}\rm\, cm^{-2}\, s^{-1}$ on June 26th, 1997.
  The 1997 mean $>1$~TeV flux level was $64\times10^{-12}\rm \,cm^{-2}\, s^{-1}$ (3.7 Crab units).
  Upper limits are given at the 2~$\sigma$ confidence level.
}
\end{figure}

Fig.\ \ref{lc} shows the diurnal averages of the Mkn~501 integral
fluxes above 1~TeV as measured during 1998 and 1999.
Compared to 1997 when diurnal averages of up to 
$\simeq178\times 10^{-12}\rm\, cm^{-2}\, s^{-1}$ (corresponding to 
10 Crab units, 1 Crab unit equals $17.5\times 10^{-12}\rm cm^{-2} s^{-1}$)
and a mean $>1$~TeV integral flux of $64\times10^{-12}\rm\, cm^{-2}\,
s^{-1}$ (3.7 Crab units) were observed, a striking decrease in flaring
activity can clearly be recognized. 
Only 1 flare with a peak flux substantially surpassing the flux level
of the Crab Nebula was detected on June 26/27, 1998 (MJD 50991) and June 27/28 (MJD 50992). 
The mean $>1$~TeV integral flux of the 1998--1999 data sample without the 1998 June
flare was $(4.8\pm0.3_{\rm stat})\times10^{-12}\rm \,cm^{-2}\, s^{-1}$, a factor of 13
lower than that of 1997.

The June 1998 flare and the correlated X-ray activity have been
discussed in detail by Sambruna et al.~(2000). 
For the day of maximum emission, June 26/27, we found evidence on the
99.4\% confidence level that the flux increased and decreased by a
factor of 2 within a time interval of 1 hour.
The spectrum of the 2 days with maximum emission seemed to be softer than that
of the 1997 observations, but the difference was not statistically significant. 
A fit of a power law with an exponential cutoff yielded the parameters (see Eq.~(\ref{plc})):
$N_0=(79\pm 10_{\rm stat}) \times 10^{-12}$ $\rm cm^{-2}\, s^{-1}\, TeV^{-1}$,   
$\alpha=1.92 \pm0.30_{\rm stat}$, and $E_0=4.1 (-0.9+1.45)_{\rm stat} \, \rm TeV$.
For the remaining days of the 1998--1999 observations we did not find 
additional evidence for
intraday variability. For the typical diurnal observation times of about 
1.5~h the 2$\sigma$ (S/N) flux sensitivity threshold was approximately 1/6th 
of the Crab flux.
For 36\% (63\%) of the days with observations,
we found an excess with more than 2$\sigma$ (1$\sigma$) statistical significance. 
For 12\% of the nights, the Mkn~501 flux was 
clearly below the sensitivity threshold for diurnal observations
and an event deficit rather than an event excess was 
found in the signal region. 

\begin{figure}[tbh]
\plotone{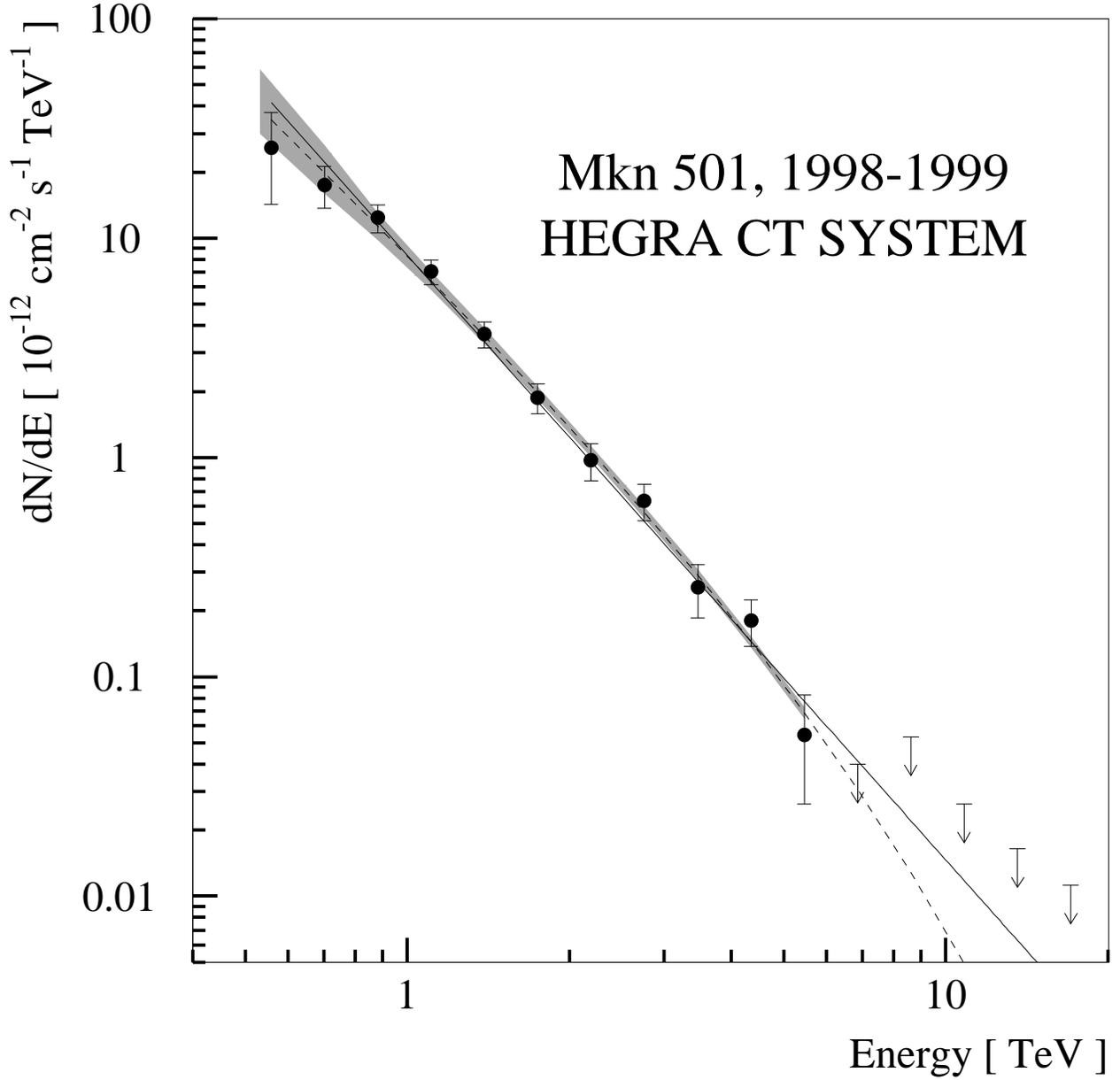}
\caption{\label{spectrum} \small The figure shows the 1998--1999
low-flux spectrum measured with the HEGRA telescope system.
The systematic error on the curvature of the spectrum is given
by the hatched region. The additional uncertainty of 15\% in 
the absolute energy scale is not shown here.
The solid and dashed lines give the fit results of a pure power law and a 
power law plus exponential cutoff model, respectively (see text).
Upper limits are given at the 2~$\sigma$ confidence level.
}
\end{figure}

We determined a ``low-flux'' TeV energy spectrum of Mkn 501 
by analyzing 122~h of 1998--1999 data, excluding the 4~h of 
data taken during the 1998-flare. 
Here we combine data of the 2 years to increase the statistical 
accuracy of the resulting spectrum. 
Analyzing the 2 years independently yields a very similar
fitted mean flux at 1~TeV for both data sets and, within the large
statistical errors, no significant difference in spectral shape.
The 1998--1999 low-flux spectrum combined with the results of the
fits discussed in the following are shown in Fig.~\ref{spectrum}.

A fit of a power law model with an exponential cut-off yields the
parameter values (see Eq.~(\ref{plc})):
$N_0=(10.1\pm 1.9_{\rm stat})
\times 10^{-12}$ $\rm cm^{-2}\, s^{-1}\, TeV^{-1}$,   
$\alpha=2.31 \pm0.22_{\rm stat}$, and 
$E_0=5.1 (-2.3+7.8)_{\rm stat} \, \rm TeV$
with a reduced $\chi^2$-value of 0.71 for 13 degrees of freedom
(the negative flux estimates corresponding to the flux upper limits 
in Fig.~\ref{spectrum} are included in the fit).
Within statistical errors, also a pure power law model 
\begin{equation}
{\rm d}N/{\rm d}E = N_0\, (E/ {\rm 1\,TeV})^{-\Gamma}
\label{pl}
\end{equation}
yields an acceptable fit with: 
$N_0 = (8.4 \pm 0.5_{\rm stat})\times 10^{-12}$ $\rm cm^{-2}\, s^{-1}\,
TeV^{-1}$, $\Gamma=2.76\pm 0.08$, and a slightly larger 
reduced $\chi^2$-value of 0.92 for 14 degrees of freedom.
The systematic error on the absolute flux is
25\%, due to uncertainty in the absolute energy scale.
Systematic uncertainty on the shape of the
spectrum is only relevant for energies below 1~TeV
and corresponds to an uncertainty of a power law spectral index of 0.05
(see the hatched area in Fig.\ \ref{spectrum}).
The results of the fits to the Mkn~501 spectra are summarized in Table \ref{t01}.

The fit results indicate that the 1998--1999 low-flux spectrum is 
softer than the 1997 time averaged spectrum.
The parameterization of Eq.~(\ref{plc}) is not suited to assess the 
statistical significance of the spectral steepening since the fit-parameters
$\alpha$ and $E_0$ are strongly correlated and hence, the statistical errors
on both parameters are large.
A fit to the ratio of the two spectra shows that indeed the spectrum 
softened significantly:
$({\rm d}N/{\rm d}E)\mbox{(1998--1999 low-flux)}~/~(
{\rm d}N/{\rm d}E)(1997)\propto E^\gamma$
with $\gamma=-0.44\pm0.10_{\rm stat}$.
The systematic error $\Delta\gamma_{\rm syst}$ 
on the change in spectral index from year to year 
is smaller than the error on the shape of individual spectra 
since several contributions to the systematic error affect all measured 
spectra in the same way; we estimate that
$\Delta\gamma_{\rm syst}$\uka~0.05, i.e.\ 
it is smaller than the statistical error on $\gamma$ of 0.1.
This latter conclusion is substantiated by observations 
of the persistent TeV emitter Crab Nebula \citep{Ahar:00}. 
The spectral index determined for the two Crab observation periods 
1997--1998 and 1998--1999 differs by only 0.02$\pm$0.07$_{\rm stat}$.
Given the statistical errors on the low-flux spectrum, it is not
possible to decide which one of the two parameters, $E_0$ or
$\alpha$, actually changed.
Fixing $E_0$ to the 1997-value of 6.2~TeV
gives a 1998--1999 photon index of $\alpha=2.36\pm0.10_{\rm stat}$, i.e.\ a value 0.44
softer than that of the 1997 energy spectrum. Fixing $\alpha$ to the 
1997-value of 1.92 gives a 1998--1999 high energy cutoff of 
$E_0=(2.61-0.38+0.44)_{\rm stat}$~TeV, i.e.\ a value reduced by 
3.6~TeV compared to the 1997 data. 

\begin{figure}[th]
\plotone{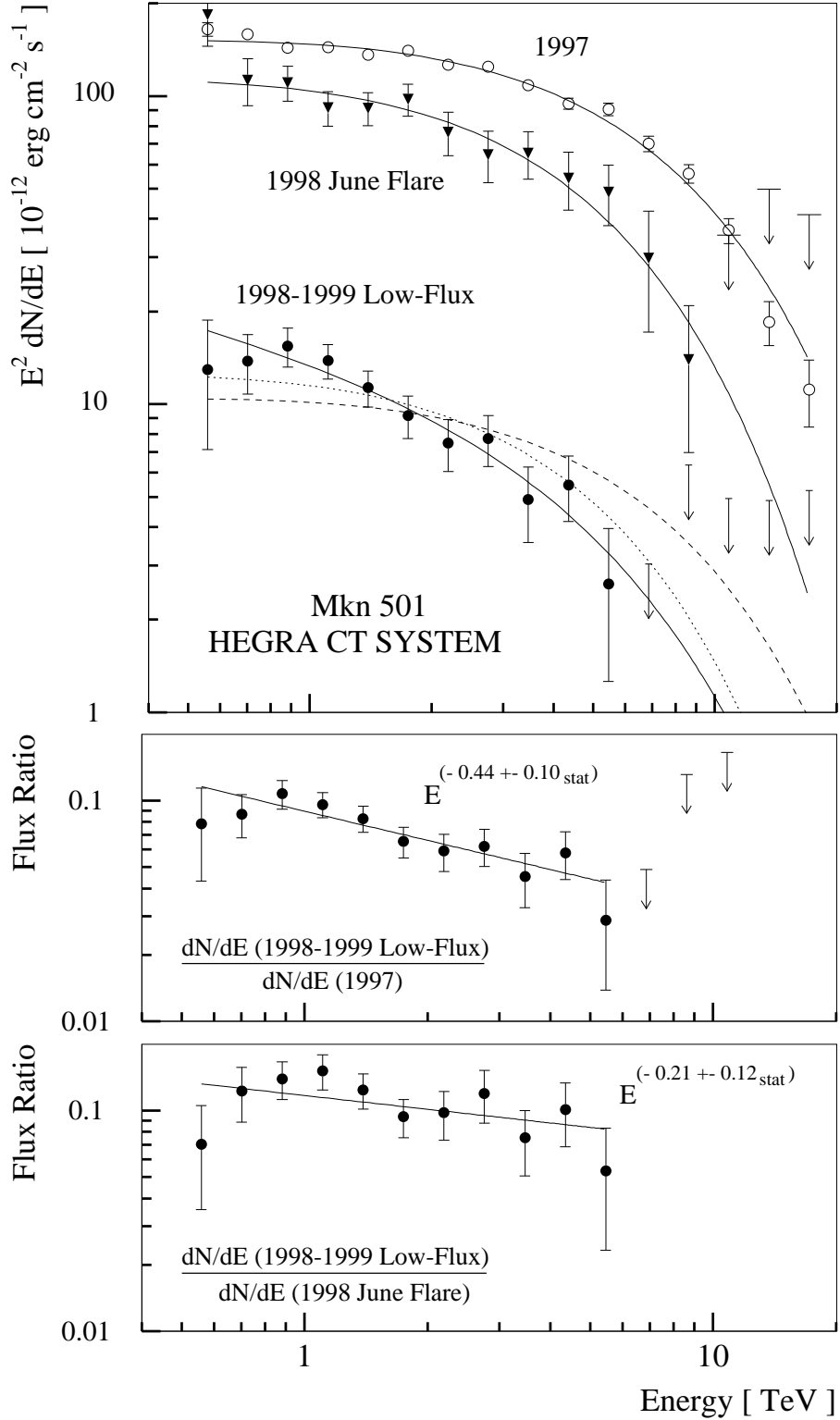}
\caption{\label{spectrum2} \small 
The upper panel shows the spectral energy distributions measured during 
1997, the 1998-flare, and the 1998--1999 low-flux period.
The solid lines show the fit results ${\rm d}N/{\rm d}E\propto
E^{-\alpha}\exp{(-E/E_0)}$ with the pairs $(\alpha,E_0)$ of
(1.9,6.2~TeV) for 1997, 
(1.9,4.0~TeV) for 1998-flare, and 
(2.3,5.1~TeV) for the 1998--1999 low-flux data sample. 
The dashed and dotted lines show the shape of the 1997 and 1998-flare 
spectra overlaid on the 1998--1999 low-flux spectrum. 
The lower 2 panels show the ratios $r(E)$ of the 1998--1999 low-flux 
and the 1997 and the 1998-flare spectrum, respectively, 
as well as the results of power law fits $r(E)\propto E^\gamma$.
All upper limits are given at the 2~$\sigma$ confidence level.
}
\end{figure}

The upper panel of Fig.~\ref{spectrum2} compares the spectral energy 
distributions as measured during 1997, during the 1998-flare,
and during the 1998--1999 low-flux phases. 
The results of the ${\rm d}N/{\rm d}E\propto E^{-\alpha}\exp{(-E/E_0)}$ fits to the data
are shown by the solid lines and the fitted spectral shapes for the 
1997 and 1998-flare data are compared to that of the 1998--1999
low-flux data by the dashed and dotted line, respectively.
The lower two panels of Fig.~\ref{spectrum2} show the ratio of
the 1998--1999 low-flux spectrum, and the 1997 and 1998-flare
spectrum, respectively.
While the 1998--1999 low-flux spectrum is significantly softer than the
1997 spectrum, its shape does not differ significantly from the
1998-flare spectrum:
the fit of a power law to the flux ratio gives:
$({\rm d}N/{\rm d}E)\mbox{(1998--1999 low-flux)}/$
$({\rm d}N/{\rm d}E)\mbox{(1998 June flare)}$ 
$\propto E^\gamma$ with $\gamma=-0.21\pm0.12_{\rm stat}$.
\section{Discussion}
\label{discussion}
In this paper we present the TeV characteristics of Mkn~501 in the
years of 1998--1999 following the major 1997 outburst phase. 
We assess for the first time the TeV energy spectrum of the source 
for a mean flux at 1~TeV well below the flux of the Crab Nebula.
This spectrum complements the high-flux spectra measured so far in the 
sense that the Mkn~501 spectrum has now been 
determined for flux levels between 1/3 Crab units (this paper) 
and $\simeq$10 Crab units \citep{Ahar:99a}.
Although the 1997 spectra were determined for a range of absolute fluxes 
differing by more than a factor of 5 (yet at 1~TeV by a factor of
at least 2 higher compared with the flux of the 1998--1999 low-flux data sample),
we did not detect any evidence for spectral variability. 
In contrast we find evidence that the 1998--1999 
low-flux spectrum is substantially softer (0.44 in spectral index)
than the 1997 spectra.

The Synchrotron Self-Compton (SSC) mechanism is widely believed to
be responsible for the non-thermal X-ray and 
TeV gamma-ray emission in BL Lac objects \cite{UUM:1997}.
A high energy population of electrons embedded in a relativistic jet
approaching the observer with a speed close to the speed of light
emits X-rays as Synchrotron radiation and 
TeV gamma-rays as Inverse Compton radiation resulting from interactions of
electrons with lower energy synchrotron photons.
The recent observations of X-ray and TeV gamma-ray flares
correlated to within less than half a day for Mkn~501
\cite{Kraw:00} as well as for the other well studied 
TeV emitting BL Lac object, Mkn~421 \cite{Mara:99,Taka:99},
are naturally explained in the SSC scenario and strongly support this
model. More detailed theoretical work is needed to decide whether the
multiwavelength data could reasonably be 
described with alternative models assuming that the non-thermal 
emission of Mkn~501 is produced by hadronic interactions of a highly 
relativistic outflow which sweeps up ambient matter \cite{Pohl:00}, 
by interactions of high energy protons with gas clouds moving across 
the jet \cite{dar:97}, or by interactions of extremely high 
energy protons with ambient photons \cite{Mann:98}, 
with the magnetic field \cite{Ahar:00b}, or with both \cite{Muec:00}.
Note that in the proton synchrotron model the stable spectral shape is 
explained by the self-regulated synchrotron cutoff, while steepening 
in the low-flux state could be explained by the drop of the acceleration rate
of protons \cite{Ahar:00b}.
While the interpretation of TeV gamma-ray spectra is hampered by the
unknown modification of the spectrum due to intergalactic extinction 
as discussed, e.g.\ by Aharonian et al. (1999b), the temporal evolution of the 
gamma-ray flux and the spectral shape is free of this uncertainty and 
should be explained by models of the origin of the TeV radiation.

Assuming for the time being that the mechanism responsible for the
X-ray and TeV gamma-ray emission has been identified, a next step 
concerns the understanding of how the emission region(s)
is (are) embedded in the jet, how an emission region evolves with time, and 
where the energy which is ultimately converted into the 
observed non-thermal X-ray and gamma-ray radiation comes from. 
Within SSC models the stability of the TeV energy spectrum during the 1997 
flares can be explained by a spectrum of accelerated electrons 
(and possibly positrons) which is stable throughout the whole flaring phase 
\cite{Ahar:99a,Kono:99b,Kraw:00}.
The steepening of the TeV energy spectrum reported in this paper can 
be accounted for, by e.g., (i)~a steepening of the
spectrum of accelerated particles streaming into the emission region;
(ii)~the shift of the break in the electron spectrum 
(caused by the synchrotron and Inverse Compton cooling of the electrons, 
or by a lower energy cutoff of the spectrum of accelerated
particles) towards lower energies; or, by (iii)~a shift of the
maximum energy of accelerated particles towards lower energies.
Detailed modeling of the now available very detailed 
multiwavelength data on Mkn~501 as given e.g.\ by Pian et al.\ (1998),
Djannati-Atai et al.\ (1999), Krawczynski et al.\ (2000), and Sambruna et al.\ (2000)
should make it possible to identify the origin of the spectral curvature observed at X-ray and
TeV energies during 1997 and to determine which properties
of the emission region(s) changed from 1997 to 1998--1999.
\acknowledgments
{\bf Acknowledgments}\\[1ex]
The support of the German Ministry for Research and
Technology BMBF and of the Spanish Research Council CYCIT is gratefully
acknowledged. GPR acknowledges receipt of a Humboldt fellowship. 
We thank the Instituto de Astrophysica de Canarias
for the use of the site and for supplying excellent working conditions at
La Palma. We gratefully acknowledge the technical support staff of the
Heidelberg, Kiel, Munich, and Yerevan Institutes. 

\clearpage

\clearpage
\begin{deluxetable}{cccccc}
\scriptsize
\tablecaption{Results of Mkn~501 spectral fits.
Statistical errors only, see text for systematic errors. \label{t01}}
\tablewidth{0pt}
\tablehead{
\colhead{Data Sample} & \colhead{$N_0\,$\tablenotemark{a,b}} & \colhead{$\Gamma\,$\tablenotemark{a}} &
\colhead{$\alpha\,$\tablenotemark{a}} &
\colhead{$E_0\,$\tablenotemark{a}} & 
\colhead{$\chi^2_r/\rm dofs$}}
\startdata
1997 & 
$108\pm2$ & --- & $1.92\pm 0.03$ & 6.2\nnerr{1.5}{2.9}&
\hspace*{0.15cm} 1.66/14 \hspace*{0.15cm} \\
June 1998 Flare & 
$79\pm10$ & --- & $\hspace*{0.15cm} 1.92\pm0.30$ \hspace*{0.15cm}&
\hspace*{0.15cm} 4.0\nnerr{0.90}{1.45}\hspace*{0.15cm} &
\hspace*{0.15cm} 0.54/13 \hspace*{0.15cm} \\
\hspace*{0.15cm}1998--1999 Low-Flux \hspace*{0.15cm}&
\hspace*{0.15cm} $8.4\pm0.5$ \hspace*{0.15cm} & 
\hspace*{0.15cm} $2.76\pm 0.08$ \hspace*{0.15cm} & --- & ---
&\hspace*{0.15cm} 0.92/14 \hspace*{0.15cm} \\
&
$10.1\pm1.9$ & --- & $2.31 \pm 0.22$ & 5.1\nnerr{2.3}{7.8} &
\hspace*{0.15cm} 0.71/13 \hspace*{0.15cm}\\
\enddata
\tablenotetext{a}{Model parameters of Eqs.\ (\ref{plc}) and
  (\ref{pl})}
\tablenotetext{b}{Flux normalization constant in ($\rm 10^{-12}\, cm^{-2}\, s^{-1}\, TeV^{-1}$)}
\end{deluxetable}
\end{document}